\documentclass[12pt]{revtex4}
\usepackage{graphicx}
\usepackage[ansinew]{inputenc}
\begin{document}
\bibliographystyle{apsrev}

\title{Universal Tunnelling Time in photonic Barriers}
\author{A. Haibel and G. Nimtz}
\affiliation{Universität zu Köln, II. Phys.~Institut, 
             Zülpicher Str.\,77, D-50937 Köln, Germany}

\begin{abstract}Tunnelling transit time for a frustrated total internal reflection 
(FTIR) in a double--prism experiment was measured using microwave radiation.
We have found that the transit time is of the same order of magnitude as the 
corresponding transit time measured either in an undersized wave\-guide (evanescent 
modes) or in a photonic lattice. Moreover we have established that in all such  
experiments the tunnelling transit time is approximately equal to the reciprocal 
($1/f$) of the corresponding frequency of radiation.
\end{abstract}

\maketitle 

Previous photonic tunnelling transit time experiments have been carried out using 
electromagnetic radiation both at microwave and optical frequencies. 
Such experiments were stimulated by the formal analogy between the classical 
Helmholtz wave equation and the quantum--mechanical Schrödinger equation. 
The corresponding tunnelling transit time data for e.g. electrons are not yet 
available.

In our Letter we are considering the tunnelling transit time for opaque photonic 
barriers~\cite{enders1,steinberg1,spielmann,ranfagni1}.
We suggest that in general the transit or delay time is approximately equal to the 
reciprocal frequency $1/f$ of the corresponding radiation and that it is independent 
of the type or shape of the actual barrier. 
The transit time or group delay time is defined as $\tau_{\rm gr}=\frac{x}{v_{\rm gr}}$, 
were $x$ is the tunnelling distance and $v_{\rm gr}= \frac{d\omega}{dk}$. 
This definition agrees with that introduced by Eisenbud and Wigner who put 
$\tau_{\varphi}= \frac{d\varphi}{d\omega} = \frac{x}{d\omega/dk}$~\cite{merzbacher}.
The tunnelling transit time or just tunnelling time for short, is measured as the 
time interval between the respective times of arrival of the signal's envelope at
the two ends of the tunnelling length $x$. 
We are not suggesting here that this is equivalent to the measure of the signal 
velocity within the barrier. 

\begin{figure}[htb]
\center{ 
\includegraphics[width=0.95\textwidth]{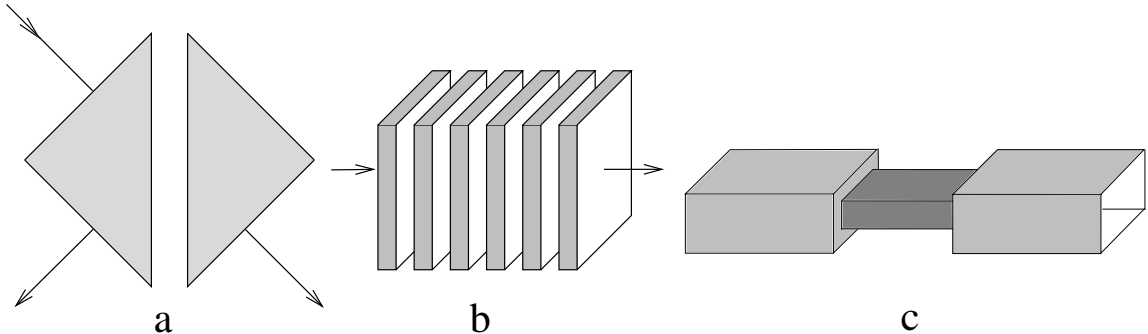}
\caption{Three types of the photonic barrier. a) A double--prism, 
         b) a photonic lattice of dielectric layers, c) an undersized waveguide 
\label{barriers}}}
\end{figure}

In order to justify this hypothesis of a universal tunnelling time we have 
carefully analyzed our own experimental results and those of others. 
Three different types of photonic barrier have been used. Investigations carried 
out in such experiments as shown in Fig.\ref{barriers}.

In Fig.\ref{barriers}a the tunnelling effect occurs between two prisms
(frustrated total internal reflection or FTIR)~\cite{carey,balcou,mugnai},  
in Fig.\ref{barriers}b tunnelling is modelled by the forbidden band of an 
one--dimensional photonic lattice~\cite{steinberg1,spielmann,nimtz1}, and in 
Fig.\ref{barriers}c tunnelling occurs in the undersized section of the 
waveguide~\cite{enders1}. Since in one dimension the Helmholtz and Schrödinger 
equations are similar, it is suggested that the three kinds of barrier can be 
used to model the one--dimensional process of wave mechanical 
tunnelling~\cite{sommerfeld,feynman}.

\begin{figure}[htb]
\center{ 
\includegraphics[width=0.35\textwidth]{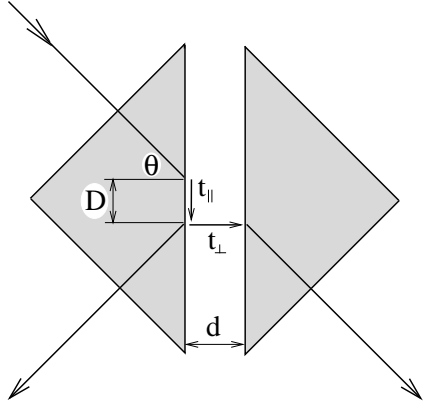}
\caption{The tunnelling time of the double-prism experiment consists of two components.
$t_{\|}$ for the Goos--Hänchen shift $D$ parallel to the prism's surface and 
$t_{\bot}$ for crossing the gap in the direction perpendicular to the two surfaces 
of the gap.
\label{doubleprism} } }  
\end{figure}

Let us start by presenting some new data on the double--prism experiment. 
For $n_1\,>\,n_2$ and an angle of incidence 
$\theta_i\,>\,\theta_c\,:=\,\arcsin\,n_2/n_1$ the incoming beam penetrates 
into the second medium and travels for some distance along the interface 
before being scattered back into the first medium (see Fig.\ref{doubleprism}); 
here $n_1$ and $n_2$ are respectively the refractive indices of the first prism 
and of the air.
If a second prism with $n_3 = n_1 = n$ is used to probe the ``evanescent'' component 
of the wave, the total reflection becomes ``frustrated'' and 
photonic tunnelling across the air gap takes place.

It is indicated in Fig.\ref{doubleprism} that the barrier traversal time of the 
double-prism, or what we call here the tunnelling time can be split into two 
components $t_{\rm tunnel}=t_{\|} + t_{\bot}$, one along the surface due to the 
Goos-Hänchen shift $D$, and another part perpendicular to the surface~\cite{artmann}. 
The measured tunnelling time represents the group or phase time delay as explained 
earlier.
The first component is related to a non-evanescent wave characterized by the real 
wavenumber $k_{\|}\,:=\,k_0\,n\,\sin \theta_i$ while the second one 
$k_{\bot}\,:=\,i\,k_0\,\sqrt{n^2\,\sin^2\,\theta_i\,-\,1}$ is related 
to the evanescent mode traversing the gap between the two prisms. 
($k_0 = 2\pi/\lambda_0$, $\lambda_0$ is the corresponding vacuum wavelength, 
and $n$ the refractive index of both prisms.)

The Goos-Hänchen shift $D$ is a sensitive function of the gap width $d$, the frequency
of radiation and its polarization, the beam width and the angle $\theta_i$ of the 
incoming beam~\cite{balcou,cowan,horowitz}.
With increasing air gap the shift reaches a constant asymptotic value 
$D$ =  $d\varphi/dk_{\|}$~\cite{balcou,cowan}, where $\varphi$ is the phase shift 
of the reflected or transmitted beam.

We have performed a double-prism experiment with two prisms of perspex, obtained from 
a 400 mm cube by a diagonal cut. The corresponding refractive index of $n = 1.605$ 
gives a total reflection angle of $\theta_c$ = 38.68$^\circ$. 
Microwave radiation at $f = 8.45$GHz generated by a 2K25 klystron was fed to a 
parabolic dish antenna which transmitted a near parallel beam to the prisms. 
(Beam spread was less than 2$^\circ$). 

In order to appoint the tunnelling time we measured the time for a signal 
travelling the closed and the opened prism. The transmission time through the 
opened prism is faster than through the closed prism. Considering the 
modifications of the path length the tunnelling time  was determinated 
from the difference of both times. 

The signal was then picked up by a microwave horn antenna and fed amplified to an 
oscilloscope (HP 54825A).
Due to the Goos--Hänchen shift (see Fig.\ref{doubleprism}) the position of the beam's 
maximum  had to be found by scanning the reflected and transmitted beams. 
It was found that the signal had a Gaussian-like shape, its half-width 
being 8~ns~\cite{nimtz2}. 

Since the total propagation time (antenna--prism--antenna) is longer than the signal
half--width, it is safe to assume that the transmitter, the prism, and the detector 
are well decoupled since there is no danger of the circuit components being coupled
by a standing wave building up. 
The experimental set--up permits asymptotic measurements. 

The tunnelling time was measured at the frequency of 8.345\,GHz (vacuum 
wavelength $\lambda_0$ = 36~mm) using a TE-polarized beam. The beam diameter was 
190\,mm and the angle of incidence was chosen to be $\theta_i$ = 45$^\circ$.
  
\begin{figure}[htb]
\center{ 
\includegraphics[width=0.5\textwidth]{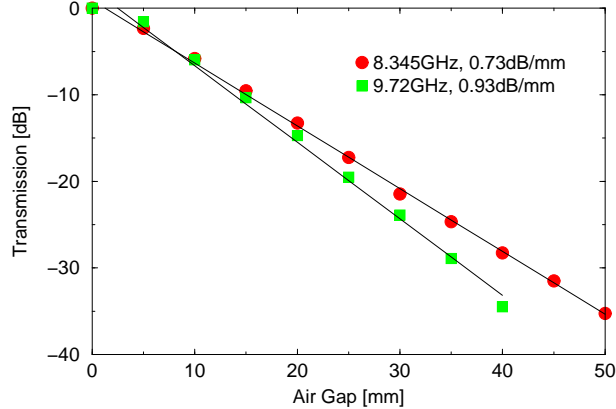}
\caption{Transmission vs Air Gap measured at two different frequencies. 
\label{transmission} } }  
\end{figure}

We tested whether all beam components were parallel and whether the angle of incidence 
was within the regime of total reflection by measuring at two different frequencies 
the transmission as a function of the gap between the two prisms. 
The measured transmission was 0.73\,dB/mm at 8.345\,GHz and 
0.93\,dB/mm for 9.72\,GHz respectively compared to the theoretical values of 
0.76\,dB/mm and 0.94\,dB/mm (see Fig.\ref{transmission}). This agreement between the 
theoretical (as quoted for $k_{\perp}$ in~\cite{feynman}) and 
experimental results indicates that our method of measuring FTIR is very sensitive, 
provided the boundary conditions are well defined.

The tunnelling time was measured in the regime of constant asymptotic Goos--Hänchen 
shift $D$, where in our case (see experimental parameters given above) $D = 31$\,mm. 
The time $t_{\rm GH} = t_{\parallel}$ for the Goos--Hänchen shift can be obtained 
from~\cite{balcou} by writing:
\begin{equation}
t_{\rm GH} \equiv t_{\|} = \frac{D\, n\,\sin(\theta_i)}{c}
\label{eq1}
\end{equation}
For $D$ = 31\,mm, n = 1.605 and $\theta_i$ = 45$^\circ$ we obtain from (\ref{eq1}) 
$t_{\rm GH}$ = 117\,ps.
Actually this value equals the measured Goos--Hänchen time for the total reflected 
beam in the absence of the second prism. (The measured time was obtained by properly 
taking into consideration the beam's path in the prism.) As mentioned before for a 
transmitted beam the total tunnelling time is the sum of the two components 
$t_{\parallel}$ and $t_{\perp}$.

Surprisingly the measured total tunnelling time proved to be equal to $t_{\parallel}$ 
alone. Since our accuracy of time measurement was $\pm$\,10 ps, this means that 
$t_{\perp} \leq$ 10\,ps or at least $t_{\perp}$\,=\,0.  

Thus it would appear that the measured total tunnelling time depends mostly on the 
Goos-Hänchen shift and hence is approximately equal to the Goos--Hänchen time 
$t_{\rm GH}$. 
This result is compatible with some theoretical investigations bearing in mind the 
imaginary wavenumber $k_{\bot}$ of the evanescent mode in the gap~\cite{stahlhofen}.

\begin{figure}[htb]
\center{ 
\includegraphics[angle=-90,width=0.8\textwidth]{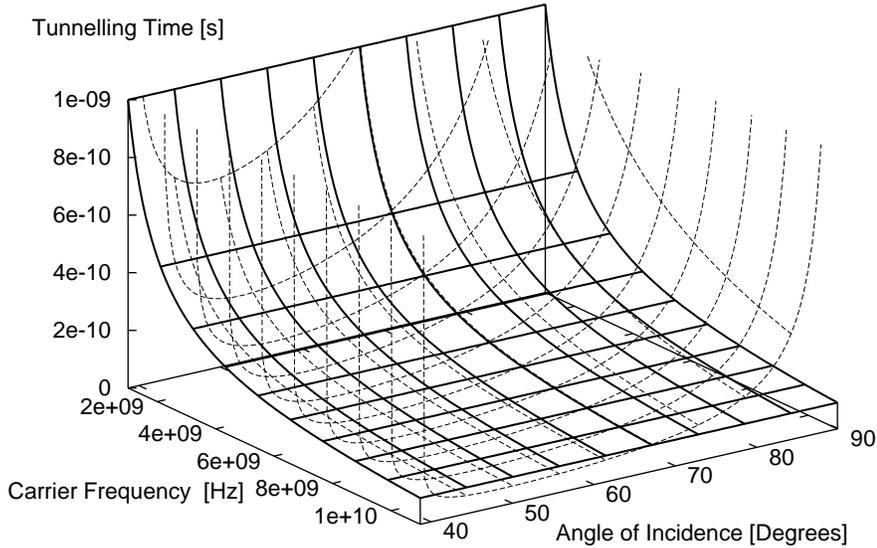}
\caption{Tunnelling time calculated using the reciprocal of the carrier frequency. 
The model of Ghatak and Banerjee~\cite{ghatak} (dashed lines) is in quite a good 
agreement with the above calculations over a wide frequency range and for most 
angles of incidence (expect the critical angle $\theta_c = 38.68^\circ$ and the 
grazing incidence (angles beyond $80^\circ$) ).
\label{ghat} } }  
\end{figure}

For large gaps where the transit time does not depend on $d$ the theoretical value 
for the FTIR-tunnelling time is 82 ps, using the model of Ghatak and 
Banerjee~\cite{ghatak}. This value is quite near to the measured value of  
117\,$\pm$\,10 ps.
It is now quite interesting to note that the reciprocal of the carrier frequency,
 $1/f=120$\,ps, gives approximately the same value for the time interval as the 
measured tunnelling time.
This result is also in agreement with the theoretical model 
of Ghatak and Banerjee, being valid over a wide range of frequencies and at all 
angles of incidence, except in the vicinity of the critical angle $\theta_c$ and 
for $\theta_i > 80^\circ$ (grazing incidence) (see Fig.\ref{ghat}). 

This relationship we have obtained for FTIR seems to be a universal property of many 
tunnelling processes.

Some previously obtained experimental results are collected in Table~\ref{table};
they all seems to confirm the suggested universal property that the tunnelling time 
is approximately equal to the reciprocal of the carrier frequency. 
Some deviations may have arisen from two experimental short comings: the studied 
barriers have not been sufficiently opaque or some tunnelling experiments were too
difficult to perform. 
\begin{table}[htb]
\begin{center}     
\begin{tabular}{|l|l|c|c|}
\hline
  Photonic Barrier & Reference  & Tunnelling Time   & Reciprocal Frequency  \\
\hline
\hline
{\it Double--Prism FTIR}   & this paper & 117\,ps & 120\,ps  \\
\cline{2-4}
  & Carey et al.~\cite{carey} & $\approx$ 1\,ps  & 3\,ps  \\
\cline{2-4}
             & Balcou/Dutriaux~\cite{balcou}  &  40\,fs &  11.3\,fs      \\
\cline{2-4}
& Mugnai et al.~\cite{mugnai} & 134\,ps & 100\,ps    \\ 
\hline
\hline
{\it Photonic Lattice} & Steinberg et al.~\cite{steinberg1} & 1.47\,fs & 2.3\,fs     \\
\cline{2-4}
&  Spielmann et al.~\cite{spielmann}  &  2.7\,fs  &  2.7\,fs     \\
\cline{2-4}
& Nimtz et al.~\cite{nimtz1}  &  81\,ps  &  115\,ps     \\
\hline
\hline
{\it Undersized Waveguide} & Enders/Nimtz~\cite{enders1} & 130\,ps & 115\,ps \\
\hline
\end{tabular}
\caption{Results of tunnelling time measurements using three different types of 
photonic barrier and  performed at quite different frequencies.  \label{table}}
\end{center}
\end{table}

Our experimental data obtained from FTIR using microwave radiation  show that the 
finite tunnelling time is largely dependent on the interference effects at the 
entrance boundary of the barrier. In the case of FTIR it is the time equivalent to 
the Goos-Hänchen shift as was first pointed out by Stahlhofen~\cite{stahlhofen}.

We have also checked using a waveguide at a microwave frequency of 8.85\,GHz, 
whether a similar behaviour applies in the case of a photonic lattice type shown 
in Fig.\ref{barriers}b. 
The measured group delay time $\tau_{\rm refl} = 75\pm5$\,ps of the reflected 
beam was found to be the same as the time measured for traversing the barrier, 
$\tau_{\rm trans} = 74\pm5$\,ps, or what we have called the tunnelling time.
Once again there is no indication that the evanescent mode spend any time inside 
the barrier, similary to FTIR~\cite{nimtz2}.

Hartman~\cite{hartman} calculated the tunnelling time (phase time delay) of Gaussian 
wave packets for one-dimensional barriers based on the time dependent Schrödinger 
equation. It is interesting to note that his theoretical wave-mechanical results 
are also in agreement with the photonic experiments for different barrier 
lengths~\cite{enders2}. 

All experimental measurements of the tunnelling time are in agreement with the 
theoretical calculations and indicate a universal tunnelling time in the case of 
opaque barriers. Both the measured finite total tunnelling time and the time delay 
of the reflected beam are associated  with the front of the barrier and closely 
correlate with the reciprocal of frequency of the corresponding radiation. 
\vspace*{1cm}

We gratefully acknowledge helpful discussions with P. Mittelstaedt, A. Stahlhofen,
R.--M. Vetter, and the Referee who made us familiar with the calculations of 
Ghatak and Banerjee. G.N. likes to thank X. Chen and P. Lindsay for stimulating 
discussions during his short stay at Queen Mary and Southfield College London.


\begin{thebibliography}{99}
\bibitem{enders1} A. Enders and G. Nimtz, J. Phys.I, France {\bf 2}, 1693 (1992) 
\bibitem{steinberg1} A. Steinberg, P. Kwiat, and R. Chiao, Phys. Rev. Letters {\bf 71}, 
                     708 (1993)
\bibitem{spielmann} Ch. Spielmann, R. Szipöcs, A. Stingle, and F. Kraus, 
                    Phys. Rev. Letters {\bf 73}, 2308 (1994)
\bibitem{ranfagni1} A. Ranfagni, P. Fabeni, G. Pazzi, and D. Mugnai, 
                    Phys. Rev. E {\bf 48}, 1453 (1994)
\bibitem{merzbacher} Merzbacher, E., {\em Quantum Mechanics}, 2nd ed., 
           John Wiley \& Sons, New York (1970) 
\bibitem{carey} J. J. Carey, J. Zawadzka, D. Jaroszynski, and K. Wynne, Phys. Rev. Letters, 
                {\bf 84}, 1431 (2000)
\bibitem{balcou} Ph. Balcou and L. Dutriaux, Phys. Rev. Letters {\bf 78}, 851 (1997)
\bibitem{mugnai} D. Mugnai, A. Ranfagni, and L. Ronchi, Phys. Letters A {\bf 247}, 281 (1998) 
\bibitem{nimtz1} G. Nimtz, A. Enders, and H. Spieker, J. Phys. I., France {\bf 4}, 565 (1994) 
\bibitem{sommerfeld} A. Sommerfeld, 'Vorlesungen über Theoretische Physik', Band IV, Optik, 
                     Dieterichsche Verlagsbuchhandlung (1950) 
\bibitem{feynman}Feynman, R. P.,Leighton, R. B. and Sands, M., 'The Feyman Lectures on 
                 Physics', Addison--Wesley Publishing Company, {\bf II} 33--12 (1964)
\bibitem{artmann} Artmann, K., Ann. Phys. {\bf 2} 87-102 (1948) 
\bibitem{cowan} Cowan, J. J. and Anicin, B., J. Opt. Soc. Am. {\bf 67} 1307-1314 (1977)
\bibitem{horowitz} Horowitz, B. R. and Tamir, T., J. Opt. Soc. Am. {\bf 61} 586-594 (1971)
\bibitem{stahlhofen} A. A. Stahlhofen, Phys. Rev. A {\bf 62}, 12112 (2000) 
\bibitem{ghatak} A. Ghatak and S. Banerjee,  Appl. Opt. {\bf 28},  1960 (1989)
\bibitem{nimtz2} G. Nimtz, Eur. Phys. J. B {\bf 7}, 523 (1999)
\bibitem{hartman} Th. Hartman, J Appl. Phys. {\bf 33}, 3427 (1962)
\bibitem{enders2} A. Enders and G. Nimtz, Phys. Rev. E {\bf 48}, 632 (1993)
%
\end{thebibliography}
\end{document}